%% file: BuildSys2015_HVACMeter.tex
\documentclass{acm_proc_article-sp}

\usepackage{graphicx}
\usepackage{balance}
\usepackage{comment}
\usepackage{times,listings,amsfonts}
\usepackage{amsmath,xcolor}
\usepackage{graphicx}
\usepackage{pifont}
\usepackage{xspace}
\usepackage{url}
\usepackage{multirow}

\def\NAME{HVACMeter\xspace} 

\begin{document}

\title{\NAME: Apportionment of HVAC Power to Thermal Zones and Air Handler Units}
%
%
%
%
%

\numberofauthors{1} 
%
\author{
\alignauthor Jason Koh$^\dag$, Bharathan Balaji$^\dag$, Rajesh Gupta$^\dag$, Yuvraj Agarwal$^\ddag$\\ 
     \vspace{2mm} 
        \affaddr{$^\dag$University of California, San Diego \hspace{10mm} $^\ddag$Carnegie Mellon University}\\
        \email{$^\dag$\{jbkoh, bbalaji, gupta\}@cs.ucsd.edu, \hspace{3mm}$^\ddag$yuvraj.agarwal@cs.cmu.edu}
}

\maketitle
\input{abstract}




\input{intro}
\input{energy}
\input{faults}
\input{discussion}

\bibliographystyle{abbrv}
\bibliography{sigproc}  
\balancecolumns
\end{document}

%% file: abstract.tex
\begin{abstract}
Heating, Ventilation and Air Conditioning (HVAC) systems consume almost half of the total energy use of commercial buildings. To optimize HVAC energy usage, it is important to understand the energy consumption of individual HVAC components at fine granularities. However, buildings typically only have aggregate building level power and thermal meters. We present \NAME, a system which leverages existing sensors in commercial HVAC systems to estimate the energy consumed by individual components of the HVAC system, as well by each thermal zone in buildings. \NAME can be generalized to any HVAC system as it uses the basic understanding of HVAC operation, heat transfer equations, and historical sensor data to estimate energy. We deploy \NAME to three buildings on our campus, to identify the set of sensors that are important for accurately disaggregating energy use at the level of each Air Handler Unit and each thermal zone within these buildings. \NAME power estimations have on an average 44.5 \% less RMSE than that of mean power estimates.  Furthermore, we highlight the usefulness of \NAME energy estimation model for a building fault detection application by quantifying the amount of energy that can be saved by fixing particular faults.   
\end{abstract}

%% file: intro.tex
\section{Introduction}
\label{sec:intro}

HVAC systems consume significant energy in commercial buildings and are naturally a focus of energy efficiency efforts. Knowing where the energy is being consumed at fine \textit{temporal} and \textit{spatial} (i.e. each thermal zone, air-handler unit (AHU), etc.) granularity is a key step towards energy optimization. Unfortunately, typical HVAC installation only provides aggregate electrical and thermal energy meters at the whole building level. While sub-metering can provide finer floor or subsystem level breakdowns \cite{agarwal2009energy}, it is costly and still not at a fine enough spatial granularity.   

Another option to estimate energy usage in different parts of the HVAC system is to use simulation software such as EnergyPlus or DOE-2. However these are only done during the design phase or during extensive retrofits given the substantial cost to build accurate models. EnergyPlus is being extended to connect to actual BMS sensor points, to provide real-time, detailed energy analysis of HVAC systems~\cite{nouidui2013bacnet}. However, these are early efforts and it takes considerable effort by an expert to build a calibrated EnergyPlus model. As a result, most building managers do not have access to calibrated energy models and rely on experience, aggregate metering, and sensors to estimate energy use at a coarse level. 

To address this challenge of energy estimation of HVAC components, we designed ZonePac~\cite{balaji2013zonepac} using a simplified energy model of VAV boxes only using heat transfer equations. ZonePAC required detailed inspection of building architectural diagrams to apply these equations, and this does not port well to HVAC systems in other buildings. In this paper, we present \NAME, a system to estimate both the heating and cooling energy of thermal zones as well as AHUs, using data driven methods to apply heat transfer equations without time consuming incorporation. We use \NAME on three buildings on our campus with installed building-level energy meters and show that we can estimate heating and cooling power with an average 44.5 \% less RMSE than that of mean thermal power.


Facilities managers struggle to maintain HVAC fault free and energy efficient~\cite{mills2010building}. On our campus, across the 55 buildings managed by a single BMS, it is common to have over 100,000 outstanding faults, and more than 10,000 fault alarms are generated \emph{per day}. While there are numerous reasons for faults being left unaddressed, the inability to easily prioritize them based on potential savings is a key limitation \cite{bdsherlock}. 
We leverage \NAME for estimation of energy wastage due to faults detected and show that it can be instrumental to help prioritize faults for facilities managers. Across three buildings we detect 19 faults with an estimated thermal energy wastage of 121 MMBTU/year.

%% file: energy.tex
\vspace{-2mm}
\section{Background}
\label{sec:hvac}

\begin{figure}[Ht]
\includegraphics[width=2.7in]{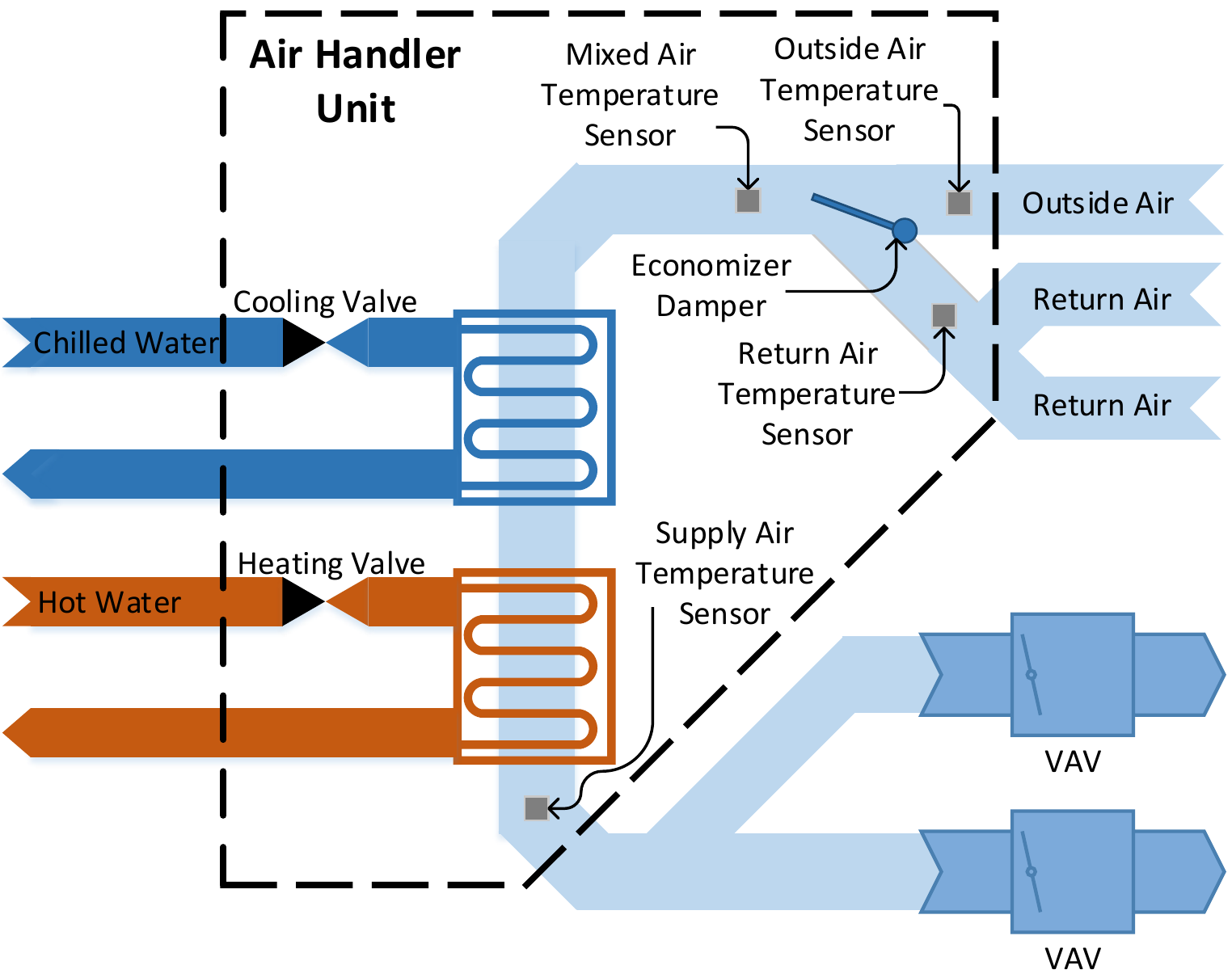}
\vspace{-3mm}
\caption{HVAC Air Handler Unit (AHU)}
\label{fig:hvac_ahu}
\vspace{-3mm}
\end{figure}

\begin{figure}[Ht]
\includegraphics[width=2.7in]{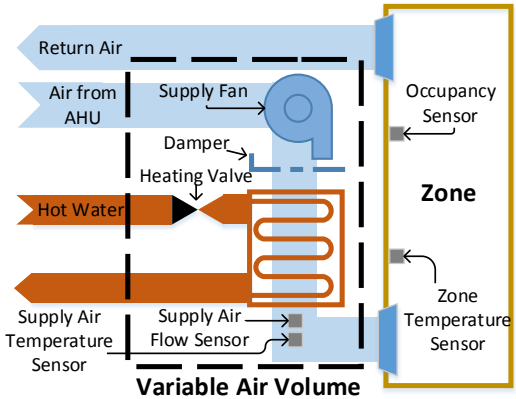}
\vspace{-3mm}
\caption{HVAC Variable Air Volume (VAV) Box}
\label{fig:hvac_vav}
\vspace{-3mm}
\end{figure}

The UC San Diego campus has a central plant that generates power and cold/hot water circulating to over 500 buildings using campus wide loops. Typical building HVAC systems contain Air Handler Units (AHUs), as shown in Figure \ref{fig:hvac_ahu}, that use this cold (hot) water, as controlled by a chilled (hot) water valve, to cool (heat) the incoming supply air as needed. The flow of air and water are maintained by fans and pumps respectively. The temperature and air flow are determined by configuration points such as \textit{supply air temperature setpoints} and \textit{supply air static pressure setpoints}. Various sensors measure temperature of supply air, return air, static pressure, etc. for closed loop control. The AHUs also mix outside air for air quality and economizing energy use by using ambient air whenever possible. For each thermal zone in a building, usually a large room or a few smaller rooms, Variable Air Volume (VAV) boxes control the environment. Often there are several hundred VAV boxes (Figure \ref{fig:hvac_vav}) in a medium sized building. Each thermal zone has a thermostat to measure temperature and allow a user to adjust settings. The amount of cold supply air in the zone is modulated using a damper, and a hot water coil can re-heat supply air when needed. In addition, some buildings may have specialized units such as dedicated chillers, exhaust fans, etc. 





\begin{figure}[Ht]
\includegraphics[width=3in]{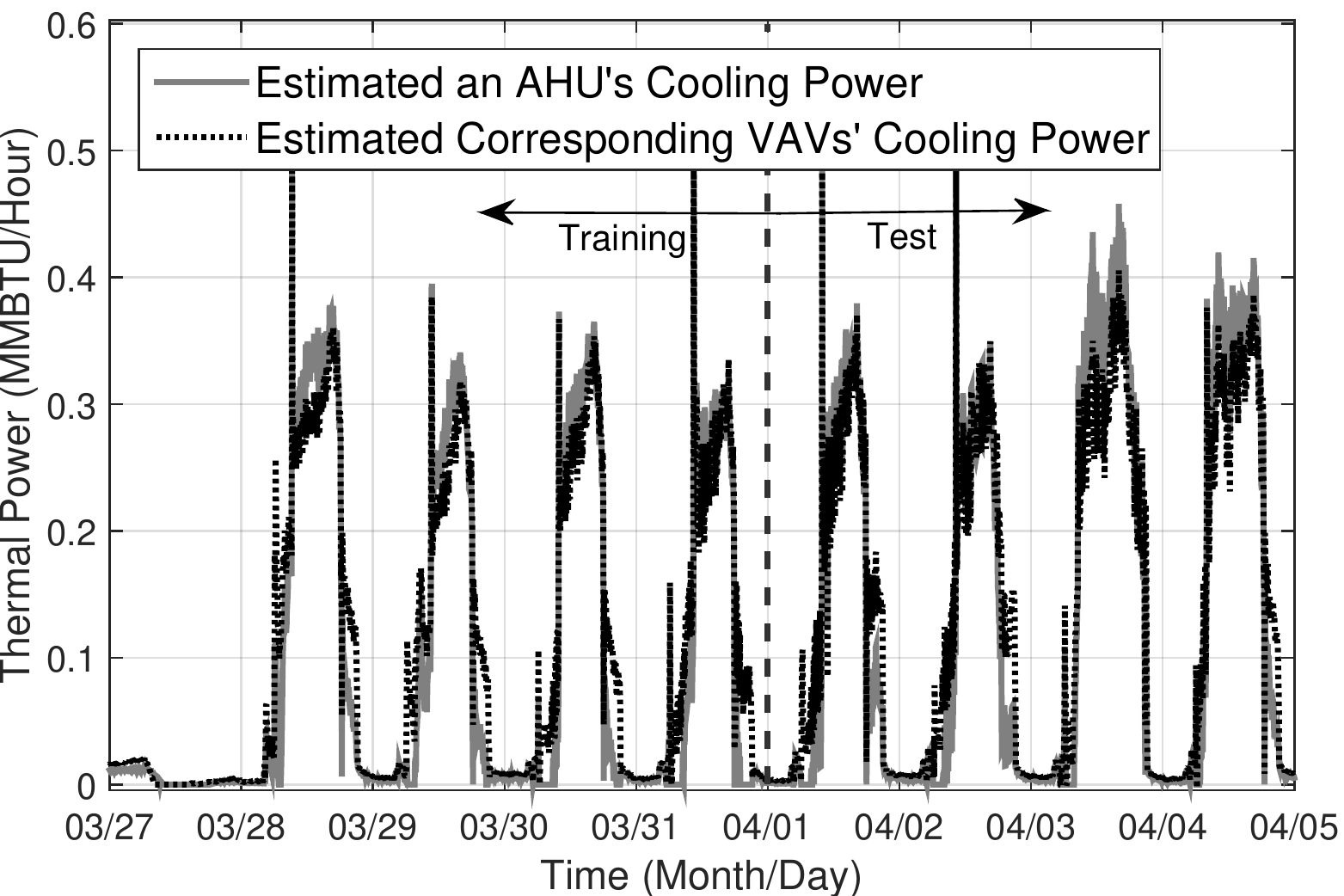}
\vspace{-3mm}
\caption{Comparison of summation of estimated VAV cooling power with estimated AHU cooling power}
\label{fig:vav_cooling}
\vspace{-3mm}
\end{figure}

\begin{figure}[Ht]
\includegraphics[width=3in]{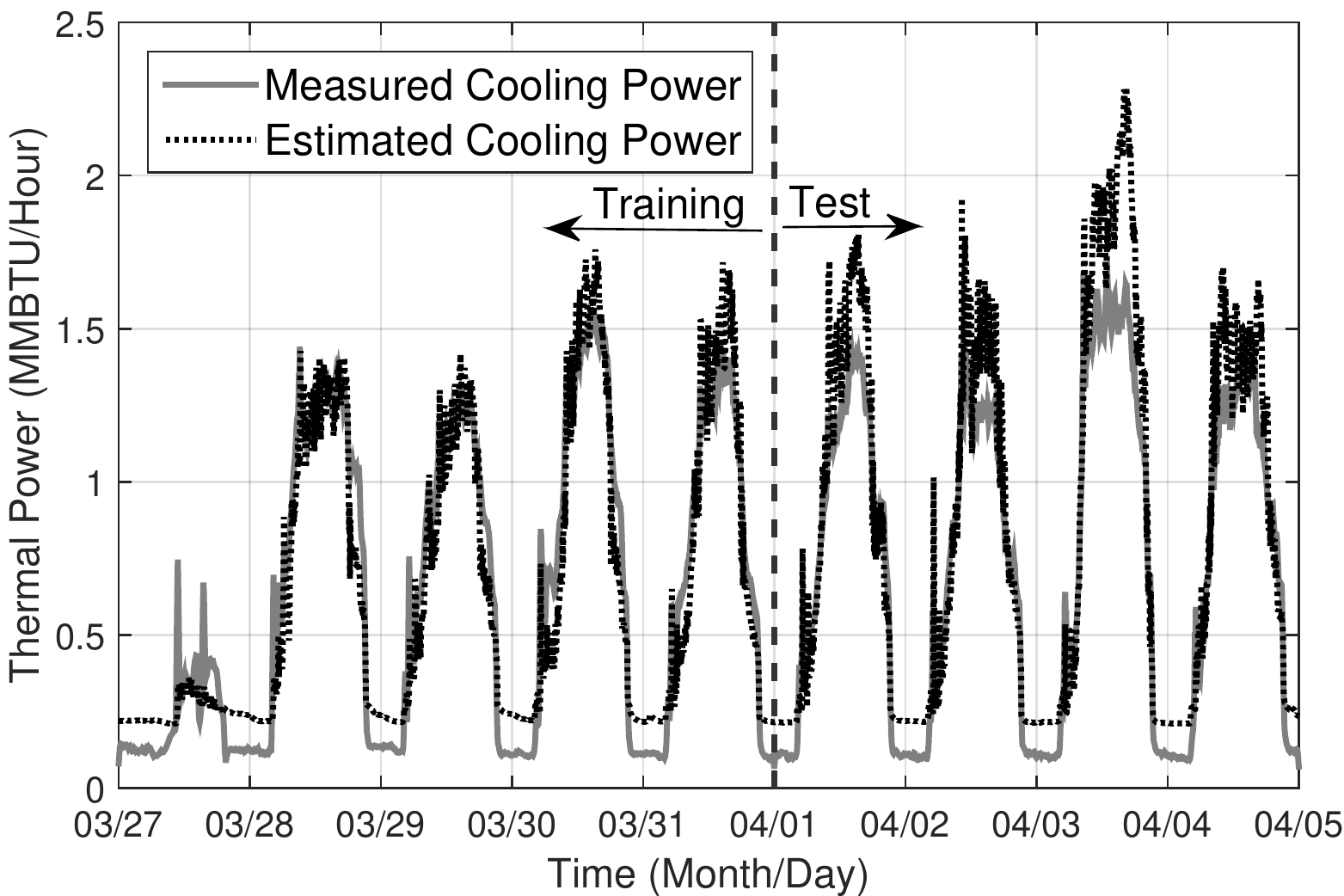}
\vspace{-3mm}
\caption{Comparison of summation of AHU cooling power with measured building cooling power}
\label{fig:ahu_cooling}
\vspace{-5mm}
\end{figure}

\begin{figure}[Ht]
\includegraphics[width=3in]{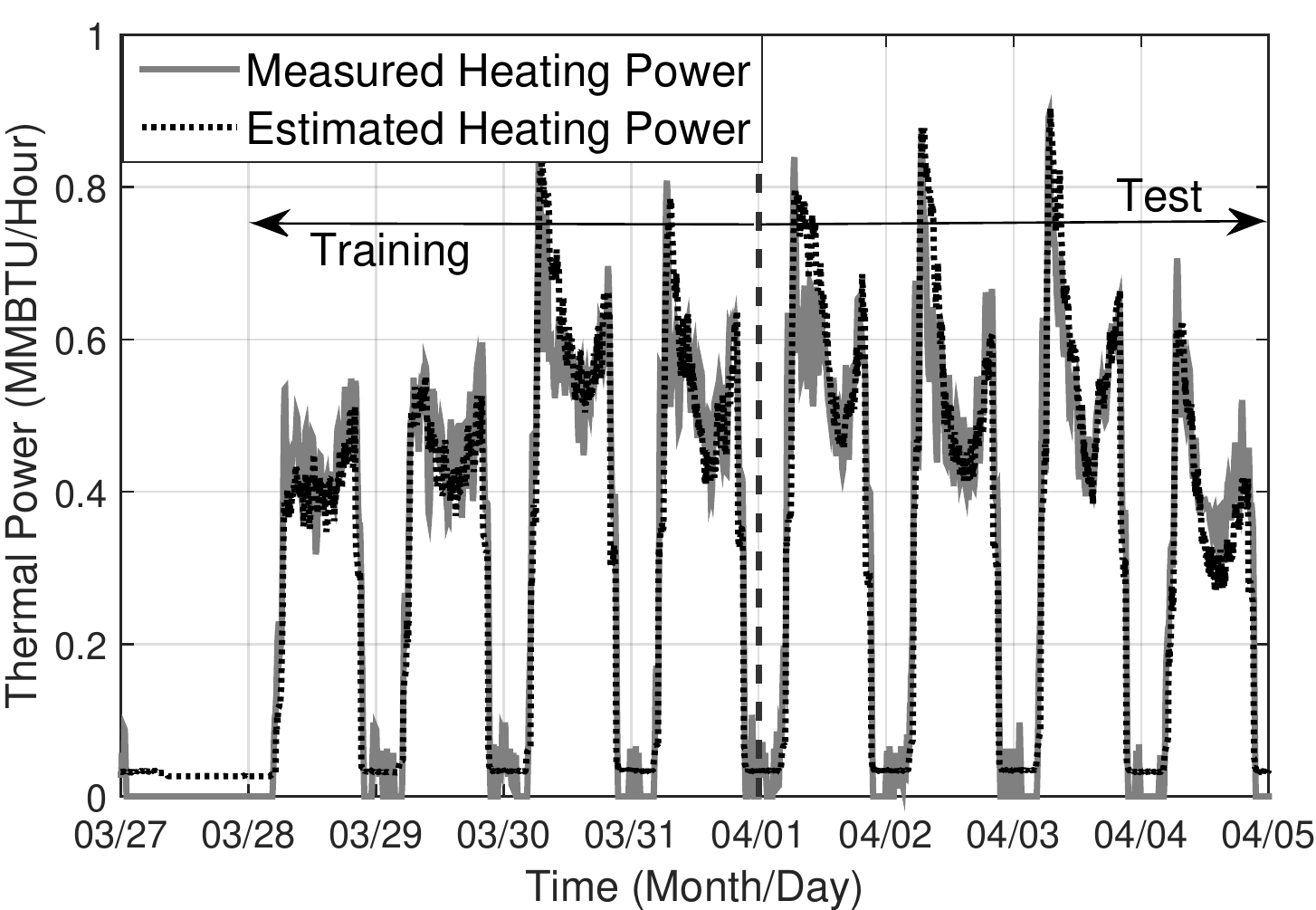}
\vspace{-3mm}
\caption{Comparison of summation of AHU and VAV heating power with building heating power measurement}
\label{fig:heating}
\vspace{-3mm}
\end{figure}

\begin{table}[Ht]
\centering
		\begin{tabular}{|p{1.1cm}| p{1cm}| p{1.5cm}| p{1cm}| p{1.5cm}|}
            \hline
            \multirow{4}{*}{\textbf{Building}} & \multicolumn{2}{c|}{\textbf{Training}} & \multicolumn{2}{c|}{\textbf{Test}}\\
							\cline{2-5}
							&\textbf{ RMSE} &  \parbox{1.5cm}{\textbf{RMSE with Mean}} & \textbf{ RMSE} & \parbox{1.5cm}{\textbf{RMSE with Mean}}\\[1.6ex]
							\hline
							& \multicolumn{4}{c|}{\textbf{Cooling Power of AHU}} \\ \hline
   		A &  0.20 & 0.83 & 0.26 & 0.34\ \\ \hline
   		B & 0.14 & 0.49 & 0.21 & 0.52 \\ \hline
		C & 0.042 & 0.093 & 0.044 & 0.090 \\ \hline
		
		& \multicolumn{4}{c|}{\textbf{Cooling Power of VAV}} \\ \hline
   		A &  0.21 & 0.83 & 0.24 & 0.34 \\\hline
   		B & 0.12 & 0.49 & 0.18 & 0.52 \\\hline
		C & 0.041 & 0.093 & 0.045 & 0.090 \\\hline
		
		& \multicolumn{4}{c|}{\textbf{Heating Power}} \\
		\hline
		A & 0.28 & 0.50 & 0.40 & 0.42 \\ \hline
		B & 0.058 & 0.25 & 0.074 & 0.15 \\ \hline
		C & 0.019 & 0.024 & 0.022 & 0.02 \\ \hline
		\multicolumn{5}{r}{(Unit: MMBTU/Year)} \\
		
        \end{tabular}
    \caption{Root Mean Squared Error of our estimation models against measured cooling and heating power as measured by whole building thermal meters versus error introduced if mean power was used as an estimate.}
    \vspace{-5mm}
    \label{tab:est_accu}
\end{table}




We rely on the basics of heat transfer for our energy estimation. According to the heat transfer equations, thermal power consumed is proportional to the rate of flow of air/water cooled and the difference between the initial and the final temperature of air/water. Hence, our goal is to estimate these physical properties from the sensors present for HVAC operation. However, some of these sensors do not directly correspond to the required rate of flow or temperature measurements, and our prior work ZonePAC~\cite{balaji2013zonepac} relied on examining building architectural diagrams to determine the relationship between the sensors and physical measurements required. Further, the heat transfer equation cannot account for errors due to sensor miscalibration, losses due to inefficiency or leakage of air/water. \NAME uses historical sensor data and data driven methods to estimate the constants that relate sensor data to the heat transfer equation variables. This principle guides the estimation models presented in the rest of the paper. We highlight the necessary sensors and the minimal HVAC structural information required for our estimation models as well as how we can compensate for lack of some of this information.

\vspace{-2mm}
\section{Cooling Power Estimation}
\label{sec:cooling}

The cooling power consumption of a VAV box is a function of the supply air temperature, the return air temperature and the amount of air circulated. It is represented by the heat transfer equation:
    \vspace{-2mm} 
	\begin{equation}
		\label{eq:heat_transfer}
		Q_{vav\_clg} = \rho*C*q*(T_{zone}-T_{supply})
	\end{equation}
where, $Q_{vav\_clg}$ = VAV cooling thermal power,
$\rho$ = density of air at $20\,^{\circ}{\rm C}$,
$C$ = specific heat of air,
$q$ = supply air flow,
$T_{zone}$ = zone temperature,
$T_{supply}$ = supply air temperature.

Since there are no return air temperature sensors in VAVs, we approximate them to the zone temperature measured by the thermostat. If the VAVs lack a supply air temperature sensor, they are approximated by the supply air temperature of the corresponding AHU. However, since HVAC systems can have multiple AHUs, knowing the mapping between a VAV and the AHU that serves it is needed. In some of the buildings we examined, some VAVs were not mapped to a thermal zone, some VAVs lacked flow sensors, and the AHU to VAV mapping was difficult to determine. We approximated these values for these rare cases.

We compare the total estimated power used by all of the VAVs with which measured for the entire HVAC system using the thermal meter. Note that the energy saved by the economizer is  estimated separately using heat transfer equations on the return air and mixed air temperatures. If mixed air temperature sensor is not available, it is estimated with outside air temperature and the economizer damper status. If return air temperature is not available, it is estimated as the average of all VAVs zone temperature. The estimated power also does not consider loss due to leakage or other factors. Hence, \NAME uses linear regression to estimate constant coefficients that better fit the total cooling power data. The final linear regression formulation follows: 

\vspace{-6mm} 
\begin{equation}
		\label{eq:vav_cooling_sum}
		C_{1}*\sum Q_{vav\_clg} + C_{2}*\sum q_{vav}*(T_{return}-T_{mixed}) + C_{3}= Q_{total}
\end{equation}
where, $q_{vav}$ = supply air flow at VAV,
$Q_{total}$ = total cooling power,
$C_{1},C_{2},C_{3}$ = regression coefficients.  As most AHUs do not have flow sensors, we sum up the air flow rate of each VAV that belongs to this AHU to obtain the total supply air flow. Figure \ref{fig:vav_cooling} compares the total estimated VAV cooling power (MMBTU/hour) with the actual value from the aggregate thermal meter. 

The estimated cooling power of an AHU is similarly a function of the supply air temperature, the mixed air temperature and the supply air flow. The summation of the cooling power of all AHUs in a building should match the measured total cooling power. We use following linear regression:
\vspace{-2mm}
\begin{equation}
		\label{eq:ahu_cooling}
		Q_{ahu} = (T_{mixed} - T_{supply})*\sum q_{vav}
\end{equation} 
\vspace{-5mm} 
\begin{equation}
		\label{eq:ahu_cooling_total}
		C_{4}*\sum Q_{ahu} + C_{5} = Q_{total}
\end{equation}
where, $Q_{ahu}$ = AHU thermal power,
$T_{mixed}$ = mixed air temperature,
$T_{supply}$ = supply air temperature,
$q_{vav}$ = supply air flow at VAV,
$C_{4},C_{5}$ = regression coefficients. When if $T_{mixed}$ $<$ $T_{supply}$, then AHU is in heating mode. We filter the data for $Q_{ahu}$ for equation \ref{eq:ahu_cooling} for AHU cooling mode only. Figure \ref{fig:ahu_cooling} compares the sum of estimated AHU cooling power with measured cooling power.

\section{Heating Power Estimation}
\label{sec:heating}
Unlike cold water, hot water is used by both VAVs and AHUs as shown in Figures \ref{fig:hvac_ahu} and \ref{fig:hvac_vav}. The heating power used by an AHU is the same as equation \ref{eq:ahu_cooling}, except that the data is now filtered for AHU heating mode and the sign has changed. For VAVs, however, there is no supply air temperature sensor available. The only points which indicate use of heating is ``heating valve command'' which controls the flow of water in the heating coil. The supply water temperature sensor is available, but the return water temperature is not, so we cannot apply the heat transfer equation as we do for AHU. We approximate the heating power at VAV as follows:
\begin{equation}
		\label{eq:vav_heating}
		Q_{vav\_htg} = (T_{supply water} - T_{supply air})*q_{vav}*H_{valve}
\end{equation}
where, $Q_{vav\_htg}$ = VAV heating power,
$T_{supply water}$ = hot water supply temperature,
$T_{supply air}$ = supply air temperature,
$H_{valve}$ = heating valve command.
This approximation is justified because VAVs heating valve command is linearly proportional to the amount of hot water used for HVAC systems we examined.

We again use linear regression as following:
\begin{equation}
		\label{eq:heating_total}
		C_{6}*\sum Q_{ahu} + C_{7}*\sum Q_{vav\_htg} + C_{8} = Q_{total}
\end{equation}
where, $C_{6},C_{7},C_{8}$ = regression coefficients. Figure \ref{fig:heating} compares total measured heating power with the sum of estimated heating power. Note that $Q_{ahu}$ is estimated using mixed air temp, and thus the savings due to economizer heating are already included. 

Table \ref{tab:est_accu} compares the estimated values with the measured power consumption. We compare the RMSE of our estimate with the RMSE if just mean power were used as an estimate. The results show not only a good match of the estimate to actual energy consumption over time, but also as summarized by Table \ref{tab:est_accu} the RMSE of \NAME is less than RMSE of mean power by 53\%, 56\%, and 27\% for AHU cooling power, VAV cooling power, and AHU+VAV heating power respectively. The variation in energy estimate across buildings is significant and subject of our continuing study. This estimation has another important use in identifying faults that we discuss next.

%% file: faults.tex
\section{Fault Prioritization}

\begin{table*}[t!]\normalsize
\vspace{-0.2cm}
\resizebox{\textwidth}{!}{
\begin{tabular}{p{1cm}p{8cm}p{4.5cm}p{2cm}p{2cm}}
 \centering {\multirow{2}{*}{\textbf{Index}}} & \centering{ \multirow{2}{*}{ \textbf{Rules}}} &  \centering {\multirow{2}{*}{ \textbf{Possible Faults}}} & \multicolumn{1}{c}{\textbf{Detected Faults}} & \multicolumn{1}{c}{\textbf{Energy Loss}}\\
\multicolumn{1}{c}{}& \multicolumn{1}{c}{} & \multicolumn{1}{c}{} & \multicolumn{1}{c}{(Number)} & \multicolumn{1}{c}{(MMBTU/Year)}\\[0.1cm]\hline\\[-0.1cm]

\parbox{1cm}{\centering 1}    & \parbox{8cm}{$Correlation(T_{MA, measured}, T_{MA, estimated}) < C_{th,1}$} & \parbox{4.5cm}{Economizer damper broken} & \parbox{2cm}{\centering 3} & \parbox{2cm}{\centering 55.56}\\[0.3cm]           

\parbox{1cm}{\centering 2}   & \parbox{8cm}{$RMSPE(q_{vav}, qs_{vav}) >C_{th,2}$ } & \parbox{4.5cm}{VAV damper leaking or stuck} & \parbox{2cm}{\centering 7} & \parbox{2cm}{\centering 49.40}\\[0.3cm]

\parbox{1cm}{\centering 3}     & \parbox{8cm}{If $unoccupied$ $\&$ $(T_{zone}<T_{upper limit})$, $q_{vav}> 1.1\cdot q_{min}$} & \parbox{4.5cm}{Configuration error} & \parbox{2cm}{\centering 7} & \parbox{2cm}{\centering 14.58}\\[0.3cm] 

\parbox{1cm}{\centering 4}     & \parbox{8cm}{When $C_{valve}=0$, $MPE(T_{SA, AHU}, T_{MA})< C_{th,3}$} & \parbox{4.5cm}{AHU cooling valve leaking} & \parbox{2cm}{\centering 2} & \parbox{2cm}{\centering 1.349}\\[0.3cm]      

               \hline\\[-0.1cm]                    




                                                                                                          


\multicolumn{5}{l}{\parbox{15cm}{
\begin{tabular}{l l l l }
T=Temperature, & qs = Supply Flow Setpoint, & SA=Supply Air, & MA = Mixed Air, \\
$C_{valve}$=Cooling Valve Command, & MPE = Mean Percentage Error, & $C_{th}$=Threshold &  q=Supply Flow, \\
AHU=Air Handler Unit, & VAV=Variable Air Volume, & RMSPE=Root Mean Square Percentage Error, & \\
\end{tabular}
}
}\\
\end{tabular}
}
		\vspace{-3mm}
    \caption{Fault Detection Rules and Result} 
    \vspace{-0.1Cm}
    \label{tab:faultrules}
    \vspace{-3mm}
\end{table*}

\begin{figure}[Ht]
\includegraphics[width=3in]{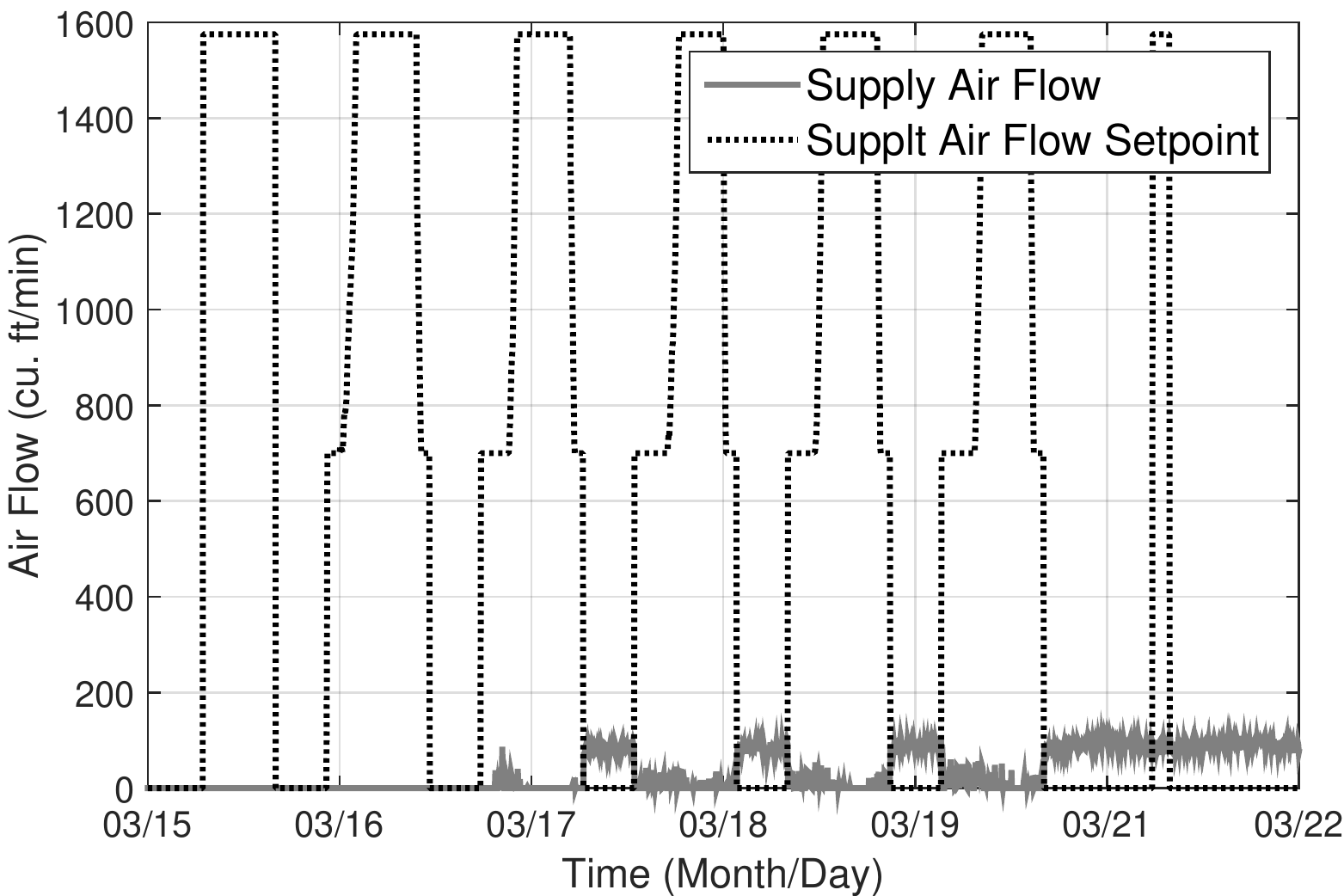}
\vspace{-3mm}
\caption{VAV fault: Air flow does not match its setpoint}
\label{fig:flow_fault}
\vspace{-3mm}
\end{figure}

\begin{figure}[Ht]
\includegraphics[width=3in]{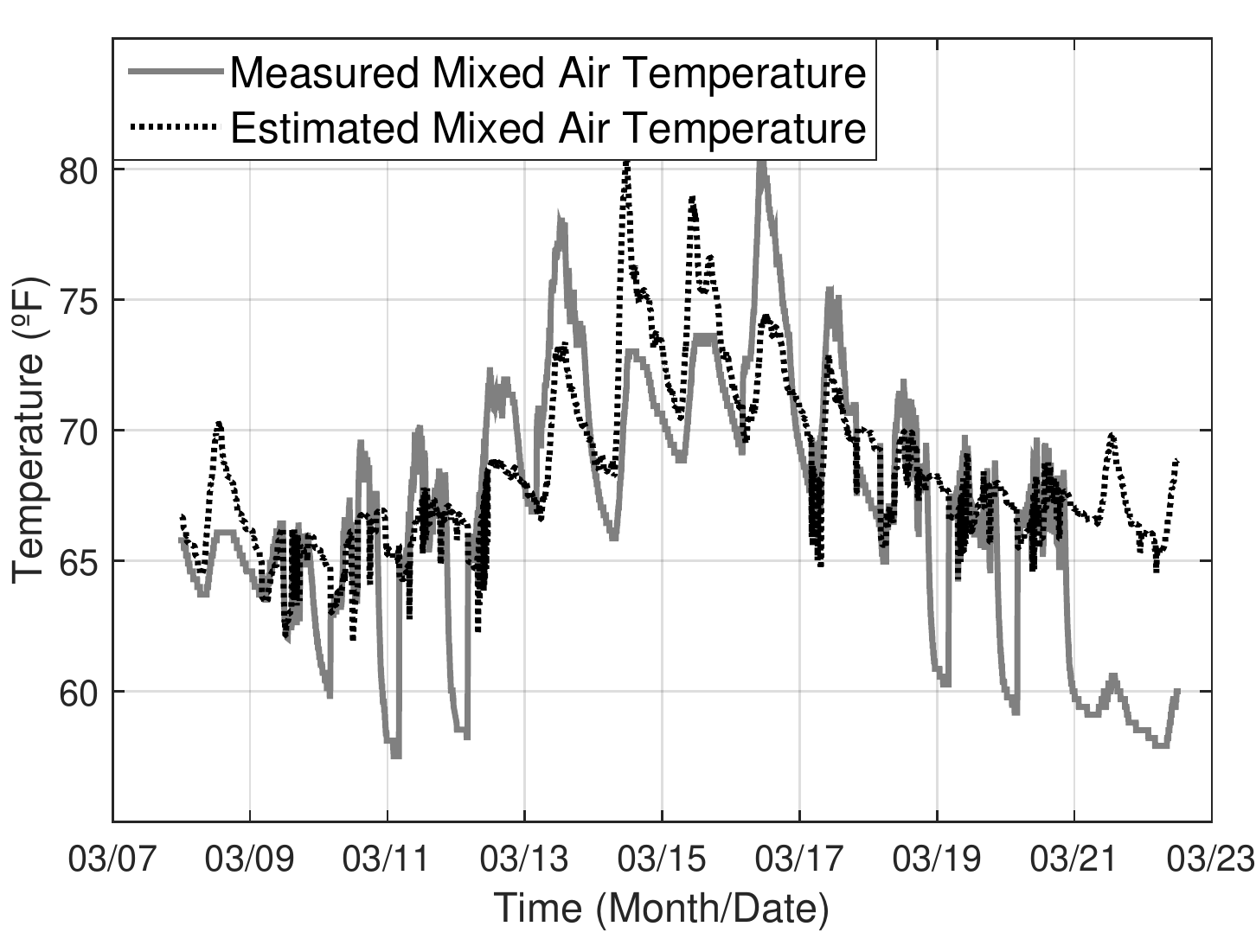}
\vspace{-5mm}
\caption{Economizer damper fault: Estimated and measured mixed air temperature do not match}
\label{fig:econ_fault}
\vspace{-5mm}
\end{figure}



We detect specific types of faults in AHUs and VAVs that are known to facility  managers, but are not detected by fault detection techniques used by BMS. We use rule based detection methods~\cite{narayanaswamy2014data,bdsherlock}, as they are supported by many fault detection frameworks~\cite{skyspark}, and can be designed to detect any type of fault. Table \ref{tab:faultrules} lists the type of faults we detect and a summary of the rules used for detection.


The generic fault detection methodology we use is to estimate the ideal values for a parameter, and compare it with its measured value. For some faults, the estimation is simple, as there points indicating the setpoint for some sensors. For example, measured supply air flow should be close to its setpoint. For other faults, we use sensor data and knowledge about HVAC working for detection. For example, we can check if the economizer damper is working correctly if the mixed air temperature changes with changes in damper position. We estimate the mixed air temperature by using measured return air temperature, outside air temperature, and economizer damper position, which determines the proportion of outside air and return air in the mixer. We can compare this estimated value to the measured mixed air temperature for detecting faults. However, it is common for temperature sensors to develop minor drifts over time, so to detect damper faults more accurately we check the correlation between the measured and estimated mixed air temperature. When they are not correlated, either one of the sensors is incorrect, or the outside air damper is damaged. Figure \ref{fig:econ_fault} shows an example of a fault detected using this metric, and we can see that the observed mixed air temperature does not change as expected with changes in economizer damper position.

To reduce false positives, and to ensure we detect faults which cause significant wastage or discomfort, we use conservative thresholds in our rules and only identify faults that persist for extended periods of time. We ignore transient and minor faults as facilities management does not have enough manpower to fix them across 55 buildings. Figure \ref{fig:flow_fault} shows an example of supply air flow fault detected using our rules, and we can observe that the air flow is almost zero even when the setpoint is set to 300 cfm. We detected total of 19 faults across three buildings (Table \ref{tab:faultrules}).

Most importantly, to prioritize the faults that cause egregious wastage, we estimate the energy impact of faults using the simplified \NAME energy model we presented in Sections \ref{sec:cooling} and \ref{sec:heating}. We estimate energy loss for those faults whose ideal values can be estimated by comparing it to the energy consumption in faulty operating conditions. We calculate savings based on at least one week of data and extrapolate the savings for a year based on correlation with outdoor weather conditions. Table \ref{tab:faultrules} summarizes the faults detected and their energy wastage. For the 19 faults detected, we estimate energy wastage of 121 MMBTU/year using \NAME. 

%% file: discussion.tex
\section{Discussion and Future Work}
We have shown that it is possible to disaggregate HVAC thermal power consumption into individual zones and AHUs. Though our estimated values are not as accurate as those provided by detailed simulation engines like EnergyPlus, \NAME provides us insight into energy flows within the HVAC system with minimal effort. With \NAME, it is easy to identify zones which consume high amount of energy and stress the AHUs, as well as identify AHUs which have abnormally high energy use due to faults, and then optimize their operation.

For data inference applications like \NAME, HVAC systems need to have a standardized ontology and historical sensor data, both of which are not common in many buildings. If they are available, innovative methods can be developed to exploit existing information to extract insights about HVAC operation. For example, AHUs across multiple buildings could be compared to check if economizers are functioning correctly. In addition, such data analysis will identify which sensors are essential for data driven monitoring of HVAC systems, and how to compensate for faulty and miscalibrated sensors.